# Outcome-Based Quality Assessment Framework for Higher Education

Samiya Khan[1], Mansaf Alam[2]

*Department of Computer Science, Jamia Millia Islamia, New Delhi*

[1]samiyashaukat@yahoo.com, [2]malam2@jmi.ac.in

**Samiya Khan** has received her Bachelor's degree in Electronics and a Master's degree in Informatics from Delhi University. She is currently pursuing her doctoral studies in Computer Science from Jamia Millia Islamia (A Central University). Her area of interest includes cloud-based big data analytics, virtualization and data-intensive computing.

**Mansaf Alam** received his doctoral degree in Computer Science from Jamia Millia Islamia, New Delhi in 2009. He is currently working as an Assistant Professor at the Department of Computer Science, Jamia Millia Islamia. He is the Editor-in-Chief, Journal of Applied Information Science. He is in Editorial Board of reputed International Journals in Computer Sciences and has published several research papers. He has also authored two books entitled "Concepts of Multimedia" and "Digital Logic design". His areas of research include Cloud database management system (CDBMS), Object Oriented Database System (OODBMS), Genetic Programming, Bioinformatics, Image Processing, Information Retrieval and Data Mining.

# Outcome-Based Quality Assessment Framework for Higher Education

This research paper proposes a quality framework for higher education that evaluates the performance of institutions on the basis of performance of outgoing students. Literature was surveyed to evaluate existing quality frameworks and develop a framework that provides insights on an unexplored dimension of quality. In order to implement and test the framework, cloud-based big data technology (BigQuery) was used with R to perform analytics. It was found that how the students fair after passing out of a course is the 'outcome' of educational process. This aspect can also be used as a quality metric for performance evaluation and management of educational organizations. However, it has not been taken into account in existing research. The lack of an integrated data collection system and rich datasets for educational intelligence applications, are some of the limitations that plague this area of research. Educational organizations are responsible for the performance of their students even after they complete their course. The inclusion of this dimension to quality assessment shall allow evaluation of educational institutions on these grounds. Assurance of this quality dimension shall boost enrolments in postgraduate and research degrees. Moreover, educational institutions will be motivated to groom students for placements or higher studies.

**Introduction**

Higher education is the backbone of the education system of any country. Effective management and quality assessment of the higher education system is not just important, but it is also necessary. However, quality assurance is usually not achieved in higher education systems because of several obstacles (Cardoso et al. 2015). The concept of quality in higher education has found varied definitions and descriptions in literature. The most recent and widely accepted definition of quality describes it as conformance of standards and meeting the set objectives. None of the dimensions covered the outcomes-based perspective of the education system.

The proposed framework evaluates quality from the perspective of how a student who passes out of an organization fairs after completion of the degree. This is assessed using the information about the university or company that student joins after course completion. Quality score is computed using this information and analytics are generated on the basis of the cumulative study of these quality scores.

The analytical framework proposed in this paper can be used for evaluating the performance of an educational organization on the basis of the cumulative quality scores' analysis of the students who pass out in a year. Moreover, predictive analysis can also be generated to monitor progress and make interventions as and when required, to maintain quality of the educational organizations and system, at large.

There are several ways in which such a quality metric may be relevant. Quality improvement approaches must be oriented with accountability (Boyle and Bowden 2006). Most quality metrics assume that the responsibility of an educational institution for a student's performance is restricted to the time that the student concerned spends enrolled in the institution. However, the responsibility of the student's performance on his or her alma mater extends beyond this timeframe. This is perhaps the reason why organizations take pride in their alumni networks and achievers who hail from their institutes. Therefore, a quality evaluation basis that assesses the organization's performance on the basis of student performance after he or she completes the course is appropriate.

Inclusion of this quality dimension to the assessment framework will bring the attention of organizations to this aspect. Quality assurance in this regard shall boost postgraduate and research enrolments, promoting higher and advanced studies. Besides this, such an assessment will also drive educational institutions to groom students for prospective future degrees or jobs, bridging the gap that usually exists between these transitions. In entirety, this quality dimension will bring educational organizations a step closer to fulfilling their purpose, meeting institutional vision and conforming to standards in the outcomes context.

In countries like India, which boast of 34,211 thousand enrolments for the academic year 2014-15 in higher education, the amount of data collected is immense (MHRD 2017). This data is high in volume, contains images and textual data for variety and is generated on a yearly basis. This makes big data technologies a relevant solution for educational analytics.

Owing to the volume, variety and velocity of data generated by the education system, education data can easily be termed as a class of big data called 'big education data'. Some of the information that is recorded as part of this system includes profile of teachers, students and operational data. Therefore, big data technologies can be used to develop educational intelligence solutions. In line with this, the case study done for implementing the proposed framework and providing a proof of concept makes use of Google BigQuery (Sato 2012) and R programming language (Siewert 2013) to generate analytics.

The rest of the paper has been organized in the following manner: Section 2.0 elaborates on the concept of quality and its definition in the higher education context. Section 3.0 provides details about the proposed framework. Section 4.0 and Section 5.0 describes how big data analytics solutions can be used for handling big data generated by higher education systems and provides a case study of how this framework can be implemented using cloud-based big data technologies like BigQuery and R programming. Lastly, the paper concludes in Section 6.0 providing insights on scope for improvements and future work.

**Concept of Quality in Education**

The standard meaning of quality is a measurement of standard or excellence. In higher education, the concept of quality emerged in early 1980s and was derived from its commercial counterparts (Newton 2002). However, academic quality was considered an abstract term in those days. The traditional concept of quality was inferred from the fact that world-class universities like Harvard and Oxford were considered benchmarks and no further dissection on the dimensions of quality was done. (Green 1994).

Green (1994) gave an extensive analysis of the literature that defines the term quality and categorized them into the following five approaches -

(1) Conforming to standards in terms of the educational process and outcomes
(2) Befitting the purpose - This is a contradictory definition as most scholars feel that if the institution meets standards, it fits its purpose, which may not always be the case.
(3) Ability to meet set institutional goals and having a clear vision
(4) Meeting the needs of the customer or student - It is important to state here that the student can be considered a product, customer or both from the higher education institution.
(5) The traditional concept which defines quality as strive for excellence

In order to elaborate on the quality dimensions for higher education, Owlia and Aspinwall (1996) gave a conceptual framework. The six dimensions identified by this framework comprise of the following –

(1) Tangibles which include infrastructure, ease of access and supporting infrastructure and facilities

(2) Competence which includes student-staff ratio and quality of staff along with their ability to communicate effectively with students

(3) Attitude which includes guidance and willingness to help

(4) Content which includes curriculum, cross-disciplinarily of knowledge and relevance of courses for future jobs

(5) Delivery which includes effective communication, student feedback and providing encouragement to students

(6) Reliability which includes matching goals and handling complaints

Another perspective that needs to be taken into account while defining quality is that of the stakeholders. In the higher education context, academics' and students are the stakeholders. However, student's perspective on quality is considered much more important (Elassy 2015). The conceptual framework (Owlia and Aspinwall 1996) caters for dimensions from a service-oriented point of view. However, the product-oriented dimensions have not been included. Related studies that cover higher education service quality dimensions include O'Neill and Palmer (2004), Telford and Masson (2005) and Angell et al. (2008).

Teeroovengadum et al. (2016) gave one of the most recent frameworks built upon the conceptual framework given by Owlia and Aspinwall (1996). This work presents an extensive Higher Educational Quality Assessment Model (HEQAM). One of the objectives of this model is career prospect of which one of the sub-objectives is perspective for professional career. However, it only covers career prospects from the perspective of tailoring the courses to meet market demands, ignoring the evaluation of the course from the perspective of how well passed out students have done in their professional careers after the completion of their courses.

**Proposed Quality Framework**

In view of the definition of quality for higher education given in the previous section, it can be understood that student is not just the customer of the system; he or she is also the product. Therefore, quality assessment and evaluation from the product perspective needs to evaluate the performance of the product. The proposed framework measures quality by computing quality scores for every student who passes out of a university in a given academic year. This score is calculated using the ranking of the organization that the student joins after passing out.

These scores are cumulatively analysed to assess the average quality score for the organization. Moreover, a year-wise analysis can provide trends and predictions in this regard. The content dimension of quality, which includes 'relevance of courses for future jobs' given by Owlia and Aspinwall (1996) is quantitatively evaluated using the quality score. Consequently, quality monitoring can be performed using analytics.

*Higher Education System – Inputs and Outputs*

The higher education system can be broken down into three academic categories namely, undergraduate, postgraduate and research. The progress pathway of a student from one academic category to the next is shown in Figure 1. Students admitted to undergraduate courses in a university, upon completion, may either choose to join a postgraduate course in the same or another university, take up a job or not pursue anything at all.

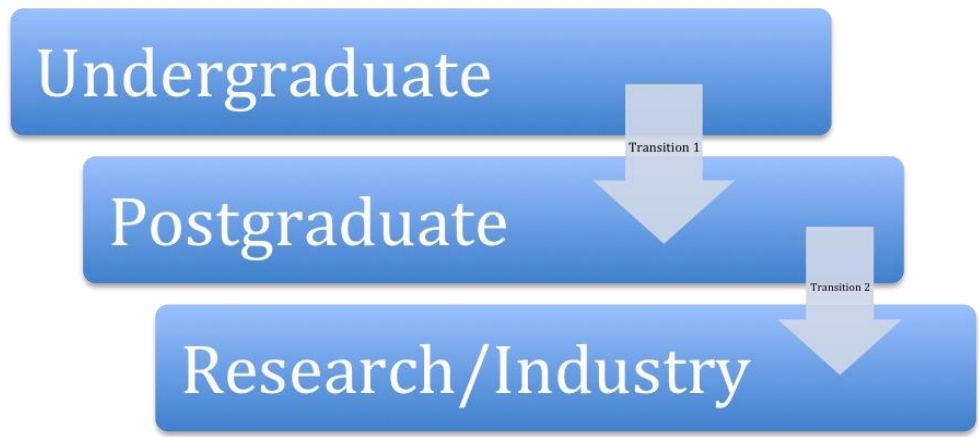

Figure 1 – Breakdown of Higher Education System

Similarly, postgraduate course students may choose to take up a job, pursue research in the same or a different university or not pursue anything at all. Evidently, there are two transitional states. The first state of transition (Transition 1) is when a student completes an undergraduate course. However, the state when a postgraduate student completes a course can be considered the second state of transition (Transition 2). This transition diagram forms the basis of the proposed quality metric and analytical framework.

From this diagram, three academic categories of students can be formulated. These academic categories include Students Pursuing Higher Education (SPHE), Students Opting For Jobs (SOFJ) and Students With No Data Available (SWNDA). SPHE includes students who choose to pursue a postgraduate degree after undergraduate degree or research after postgraduate degree. SOFJ includes students who opt for on-campus placements or get an off-campus placement in the following academic year. Therefore, all students who are able to find an industry position within one year of leaving college are considered under this academic category.

The last academic category, SWNDA, includes students who do not fall under the other two academic categories. Since, this is an yearly analysis, any student who finds an off-campus placement after one year, chooses self-employment or does nothing at all for the first year after graduation, is considered under this academic category. This academic category has been added for comprehensiveness. However, for simplicity sake, the quality score for this academic category of students is taken as zero.

Understandably, self-employment is a special case scenario. However, even if the student takes up self-employment within the first year of leaving the university or institution, there is no parameter to judge the success of his or her venture in the given time period. Therefore, a more detailed framework shall be required for calculating quality score for the self-employed.

### *Quality Score (QS) - Metric for Quality Assessment*

This research paper proposes a metric termed as 'quality score', which shall be calculated at student level on the basis of the transition outcome of the student from one state to another. The quality scores for students enrolled to a university shall be calculated year-wise on a scale of 1 to 10 and used for further analytics. Quality score calculations for SPHE and SOFJ academic categories of students have been provided in the following sections.

### *Quality Score Calculation for SPHE*

As mentioned previously, students, upon completion of their undergraduate or postgraduate degrees are expected to pursue higher education or opt for campus placements. University or institute rankings are provided by Government organizations and private ranking agencies on a yearly basis.

In order to ensure and maintain authenticity of the base data used for analysis, ranking provided by Government agencies is recommended for SPHE. Moreover, a student may move to a university in the same country or may opt to study abroad. In order to accommodate this case, the data for world university ranking must be taken for score calculation if the student is taking admission abroad and country-wise ranking shall apply in the other case to accommodate for maximum universities.

In order to calculate QS for every student, the rank of the university in which the student is taking up a postgraduate degree or pursuing research and the maximum rank assigned by the ranking to any university are required as inputs. If the university in which the student is taking up postgraduate degree or research is not ranked in the list, the value zero is assigned to quality score for that student.

The value of QS for a SPHE student with a known university rank is calculated by performing linear scaling. The formulae used for linear scaling (PennState 2017) are as follows –

$$rate = \frac{scaled_{max} - scaled_{min}}{input_{max} - input_{min}} \qquad (1)$$

$$offset = scaled_{min} - (input_{min} * rate) \qquad (2)$$

$$ouput = (input * rate) + offset \qquad (3)$$

The derivation of the formula for QS calculation in this scenario is given below. The value of variables used in equations (1), (2) and (3) are as follows –

$input = rank$

$output = QS_{SPHE}$

$input_{min} = rank_{max}$

$input_{max} = 1$

$$scaled_{min} = 1$$

$$scaled_{max} = 10$$

Substituting these values in equations (1), (2) and (3),

$$rate = \frac{9}{1 - rank_{max}} \quad (4)$$

$$offset = 1 - (rank_{max} * rate) \quad (5)$$

$$QS_{SPHE} = (rank * rate) + offset \quad (6)$$

A description of the variables used in the described formula is given in Table 1.

Table 1 – Variables used for SPHE Quality Score Calculation

| Variable | Description |
| --- | --- |
| $input$ | The input value that needs to be scaled |
| $output$ | The scaled value |
| $input_{min}$ | The minimum value of the input scale |
| $input_{max}$ | The maximum value of the input scale |
| $scaled_{min}$ | The minimum value of the output scale |
| $scaled_{max}$ | The maximum value of the output scale |
| $rate$ | The rate of scaling |
| $offset$ | The offset that needs to be applied for scaling |
| $rank$ | The university rank for the concerned SPHE student |
| $rank_{max}$ | The maximum rank assigned to a university in the University Ranking data used |
| $QS_{SPHE}$ | Quality Score for SPHE of the concerned student |

*Quality Score Calculation for SOFJ*

Placement data like company ranking and package offered can be cumulatively used as base data for students who opt for campus placed jobs. The company rankings can be taken from survey results from credible private agencies like Economic Times (Times Internet Limited 2015), which have dedicated surveys with the objective to create top recruiters list. Data for package offered by companies to individual students is available with the university and can be directly used for analysis.

Given the fact that the best student performer in this academic category is the one who gets placed in a company with highest ranking and gets the highest package. On the other hand, the worst performer is the one who gets placed in an unranked company at the lowest package. In order to compute the total quality score for SOFJ, QS calculated on the based of industry rankings is added to QS calculated on the basis of relative package score.

Quality score computed on the basis of industry ranking ($QS_{IR}$) makes use of the same concept as used by $QS_{SPHE}$. If the company in which the student is taking a campus placement is not ranked in the list, the value zero is assigned to for $QS_{IR}$ of that student. In order to accommodate this case, the value of $QS_{IR}$ is calculated on a scale of 1 to 5. The value of $QS_{IR}$ can be calculated using the equations (1), (2) and (3) with the following parametric values:

$input = industry\_rank$

$output = QS_{IR}$

$input_{min} = industry\_rank_{max}$

$$input_{max} = 1$$

$$scaled_{min} = 1$$

$$scaled_{max} = 5$$

Substituting these values in equations (1), (2) and (3),

$$rate = \frac{4}{1 - industry\_rank_{max}} \tag{7}$$

$$offset = 1 - (industry\_rank_{max} * rate) \tag{8}$$

$$QS_{IR} = (industry\_rank * rate) + offset \tag{9}$$

Scaling of package offered to students opting for campus-placed jobs on a scale of 1 to 5 also requires linear scaling. Therefore, equations (1), (2) and (3) are used with the following parameters:

$$input = package$$

$$output = QS_{PO}$$

$$input_{min} = package_{min}$$

$$input_{max} = package_{max}$$

$$scaled_{min} = 0$$

$$scaled_{max} = 5$$

Substituting these values in equations (1), (2) and (3),

$$rate = \frac{5}{package_{max} - package_{min}} \tag{10}$$

$$offset = -(package_{min} * rate) \tag{11}$$

$$QS_{PO} = (package * rate) + offset \tag{12}$$

Quality score for SOFJ ($QS_{SPHE}$) is determined using the following equation,

$$QS_{SOFJ} = QS_{IR} + QS_{PO} \tag{13}$$

A description of the variables used in the described formula is given in Table 2.

Table 2 – Variables used for SOFJ Quality Score Calculation

| Variable | Description |
| --- | --- |
| $input$ | The input value that needs to be scaled |
| $output$ | The scaled value |
| $input_{min}$ | The minimum value of the input scale |
| $input_{max}$ | The maximum value of the input scale |
| $scaled_{min}$ | The minimum value of the output scale |
| $scaled_{max}$ | The maximum value of the output scale |
| $rate$ | The rate of scaling |
| $offset$ | The offset that needs to be applied for scaling |
| $industry\_rank$ | The company rank for the concerned SOFJ student |
| $industry\_rank_{max}$ | The maximum rank assigned to a company in the Company Ranking data used |
| $QS_{IR}$ | Quality Score for SOFJ of the concerned student based on company ranking |
| $package_{min}$ | The minimum package that has been offered to any SOFJ student |
| $package_{max}$ | The maximum package that has been offered to any SOFJ student |
| $package$ | Package offered to the concerned SOFJ |

|  | student |
| --- | --- |
| $QS_{PO}$ | Quality Score for SOFJ of the concerned student based on package offered |
| $QS_{SOFJ}$ | Cumulative Quality Score for SOFJ of the concerned student |

*QS-Based Analytics for Higher Education Institutions*

Once quality score is determined for the students passing out of an institute or university at the end of a specific academic year, these values can be used to generate varied types of analytics and graphical interpretations. In order to use QS for decision-making and organizational efficiency management, students need to be divided into three QS categories depending on the QS calculated for the student. Figure 2 illustrates the framework that needs to be used for student categorization.

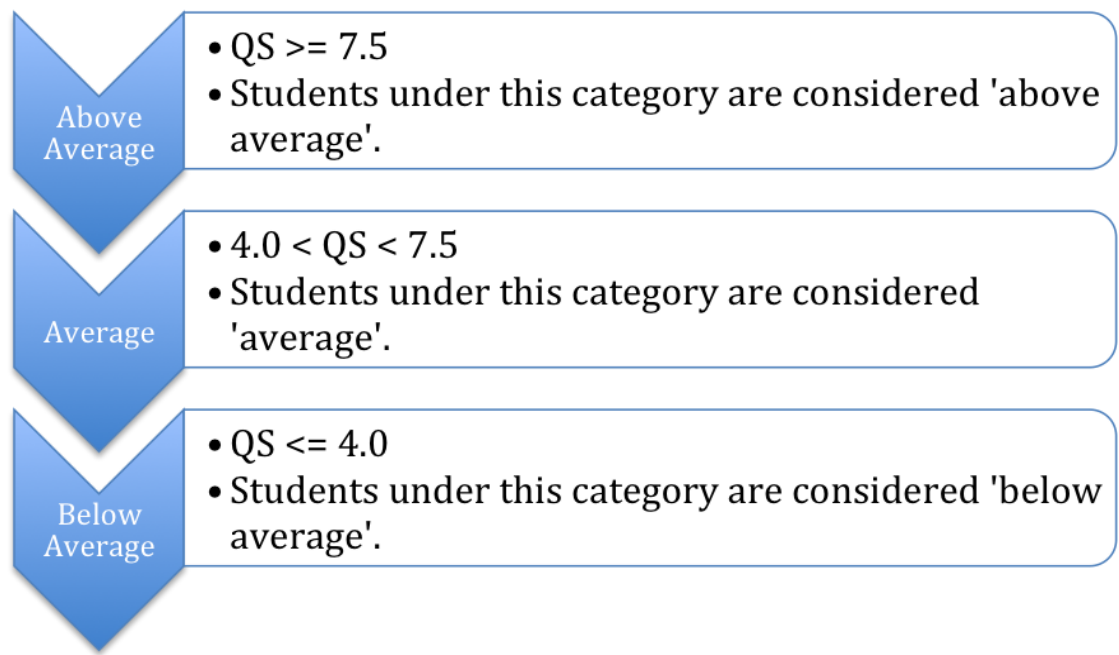

Figure 2 – QS-based Quality Framework

On the basis of this categorization, average QS for each QS category can be determined. Moreover, graphical illustrations like pie charts indicating the share of 'Above

Average', 'Average' and 'Below Average' to indicate organization performance for a specific year can be created. Besides this, analytics that use data generated over the years can be used to create line charts for indicating the performance patterns of the organization in all the three categories. These analytics can be used to assess the performance of an institute or organization for an academic year as well as over the years. Cumulatively, these visualizations can be used for performance assessment and comparison.

**Methodology and Implementation**

The applicability of big data technologies, implemented using cloud-based infrastructures, to real-world, data-intensive scenarios has given rise to many sub-research areas. Education data is a form of big data and can make use of big data technologies for generating valuable analytics for an educational organization (Daniel 2017).

IBM's big data model defines five big data characteristics namely, volume, velocity, variety, veracity and value (Assunção et al. 2015). Evidently, data collected at the student level is 'big' in 'volume', both on a per-year basis as well as over-the-years. Student profiles include textual and image data like scanned copy of the student's photograph and signature. The different types of data included in the collection makes up for the 'variety' characteristic of big data. Moreover, this data is generated on a yearly basis, accounting for 'velocity' of data.

Student data used for processing is manually entered. As a result, the probability of error and associated uncertainty are rather high, which makes veracity a significant characteristic in the education system context. Lastly, educational intelligence solutions can be used for improving the operational efficiency of the system and support administrative processes with improved decision-making, adding value to the data and analytical solutions that it can produce (Khan et al. 2016).

For the purpose of feasibility evaluation of educational intelligence solutions for real world education systems, the Indian higher education system has been considered. The numbers of universities and institutions that can utilize educational intelligence solutions have been given in Table 3, with a breakdown of the different types of organizations that it entails.

Table 3 - Number of Higher Education Institutions by Type 2014-15 (MHRD 2017)

| **Level** | **Type** | **Total** |
|---|---|---|
| **University** | Central University | 43 |
| | State Public University | 316 |
| | Deemed University | 122 |
| | State Private University | 181 |
| | Central Open University | 1 |
| | State Open University | 13 |
| | Institution of National Importance | 75 |
| | State Private Open University | 1 |
| | Institutions Under State Legislature Act | 5 |
| | Others | 3 |
| | **Total** | **760** |
| **College** | | **38498** |
| **Standalone Institutions** | Diploma Level Technical | 3845 |
| | PGDM | 431 |
| | Diploma Level Nursing | 3114 |
| | Diploma Level Teacher Training | 4730 |
| | Institute Under Ministries | 156 |
| | **Total** | **12276** |

Besides this, the total number of students pursuing different courses in the Indian higher education system has been detailed in Table 4.

Table 4 - Level-wise Enrolment in School & Higher Education 2014-15 (MHRD 2017)

| Level | Total (in thousand) |
|---|---|
| Ph.D. | 118 |
| M. Phil. | 33 |
| Postgraduate | 3853 |
| Undergraduate | 27172 |
| PG Diploma | 215 |
| Diploma | 2508 |
| Certificate | 170 |
| Integrated | 142 |
| **Higher Education Total** | **34211** |

The proposed framework utilizes student data profiles to compute quality score per student and use the computed values for advanced analytics. Profiles of students passing out each year are scanned for computing quality scores and quality score data per year is stored for generating time-based analytics. Evidently, the numbers of institutions that can utilize such solutions are high. This makes educational intelligence solutions commercially viable. Therefore, big data technologies are deemed most appropriate for developing analytical solutions for this domain.

In order to implement and test the proposed framework, Google BigQuery (Google Cloud Platform 2017) and R (The R Foundation 2017) are used as based technologies. A web-based application was developed using shiny package (Chang et al. 2017) available for R. The backend programming for computation of quality score and creation of visualizations like pie charts and line charts was performed using R programming language.

The dataset used for computation was stored in BigQuery, which is a cloud-based big data-warehousing technology. The schema for the three tables used for the implementation of this framework has been shown in Table 5, Table 6 and Table 7.

Table 5 – Table Schema for Student Data Table

| Field Name | Data Type | Description |
| --- | --- | --- |
| course | STRING | Course to which the student is enrolled |
| eyear | INTEGER | Year of enrolment |
| code | STRING | Code of course to which the student is enrolled |
| id | INTEGER | Student ID |
| gender | STRING | Gender |
| region | STRING | Region to which the student belongs |
| he | STRING | Highest education |
| imd | STRING | IMD Band |
| age | STRING | Age Bracket to which the student belongs |
| prev_attempt | STRING | Number of previous attempts taken |
| credit | STRING | Credits studied |
| disability | STRING | Whether suffering from a disability |
| final_result | STRING | Final result (Pass or Fail or Withdrawn) |
| univ | STRING | University to which student has taken admission after course completion |
| comp | STRING | Company that the student has joined after course completion |
| package | FLOAT | Package offered |
| univ_f | STRING | Whether joined a university after course completion (Y/N) |
| comp_f | STRING | Whether joined a company after course completion (Y/N) |
| q_score | FLOAT | Quality score (assigned to zero for initialization) |

Table 6 – Table Schema for University Rank Table

| Field Name | Data Type | Description |
| --- | --- | --- |
| univ_code | STRING | University code |
| univ_name | STRING | Name of the university |
| univ_city | STRING | City of location |
| univ_state | STRING | State of location |
| univ_score | FLOAT | Score |

| Field Name | Data Type | Description |
| --- | --- | --- |
| univ_rank | INTEGER | Rank |
| uryear | INTEGER | Year in which rank was generated |

Table 7 – Table Schema for Company Rank Table

| Field Name | Data Type | Description |
| --- | --- | --- |
| comp_name | STRING | Name of the company |
| comp_sector | STRING | Sector of operation |
| comp_subsector | STRING | Sub-sector of operation |
| comp_area | STRING | Area/Continent of operation |
| comp_country | STRING | Country of operation |
| comp_para1 | FLOAT | Financial parameter |
| comp_para2 | FLOAT | Financial parameter |
| comp_para3 | FLOAT | Financial parameter |
| comp_para4 | FLOAT | Financial parameter |
| comp_rank | INTEGER | Rank |
| cryear | INTEGER | Year for which rank was generated |

The choice of these technologies amongst a plethora of available big data technologies and tools was made because of the cost-effective nature of BigQuery and the simplicity of R language for developing analytical visualizations. Moreover, the availability of BigQuery API, which can be easily integrated with R programming environment is also one of the reasons for this selection. The developed tool is named "Quality Management Tool for Higher Education Systems". The user interface of the web application is shown in Figure 3.

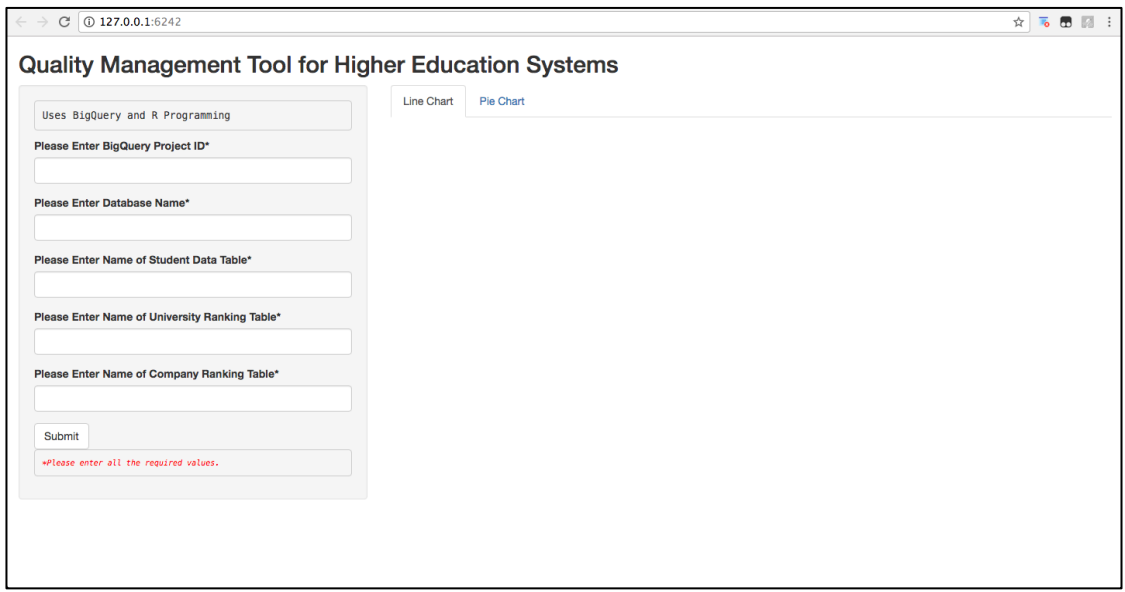

Figure 3 – User Interface of 'Quality Management Tool for Higher Education Systems'

**Case Study**

The dataset used for testing the implementation of this framework contains dummy data for students, university rankings and company rankings for the years 2013 and 2014. Quality score for every student is calculated on the basis of whether the student joins a research degree or gets a job in a company. The rank of the university determines the quality score of students joining a university. On the other hand, company ranking and package offered determine the quality score of students joining a company. The computation of quality score for the dataset used has been illustrated in Figure 4 and Figure 5.

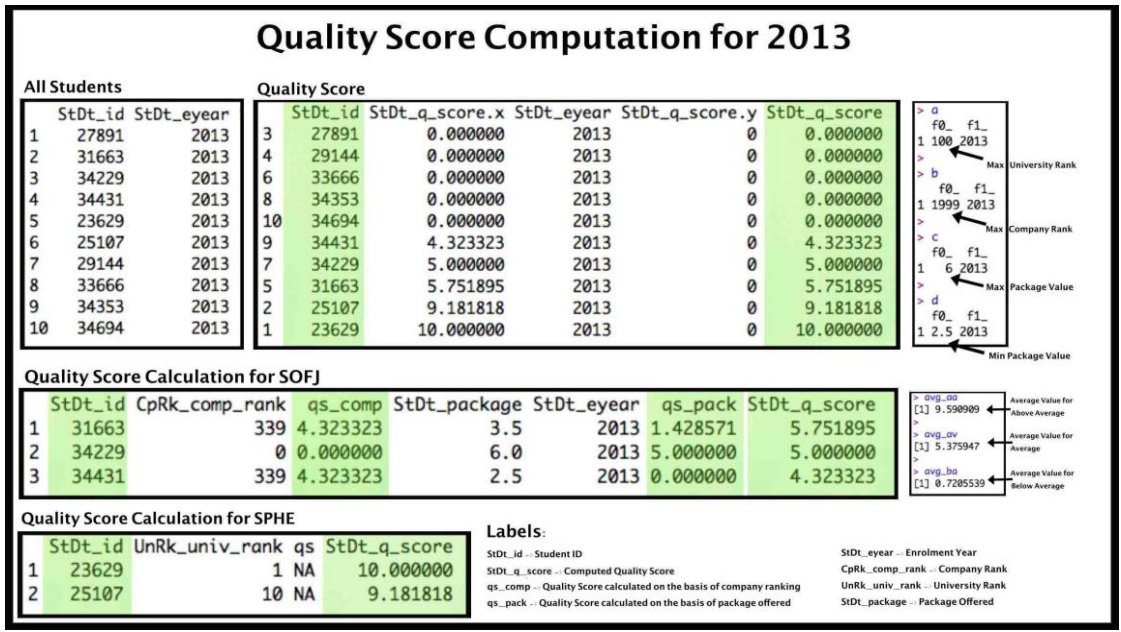

Figure 4 – Computation of Quality Scores for Students enrolled in the Year 2013

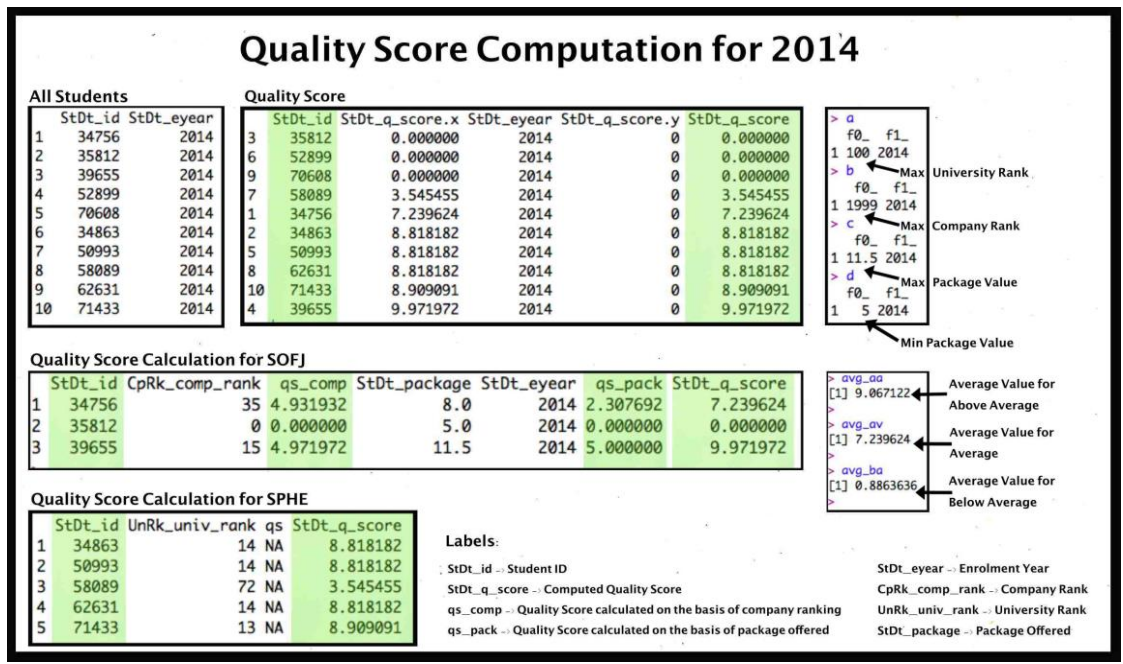

Figure 5 – Computation of Quality Scores for Students enrolled in the Year 2014

On the basis of the quality scores generated for every student passing out of the university during one academic year, average quality scores are calculated. These average values are used to generate line charts for demonstrating trends of quality score values for an educational organization. On the other hand, the numbers of students falling under specific QS categories are used to generate piecharts for specific years.

The analytics – pie charts and line chart – created using quality score data computed for the dataset used are shown in Figure 6, Figure 7 and Figure 8.

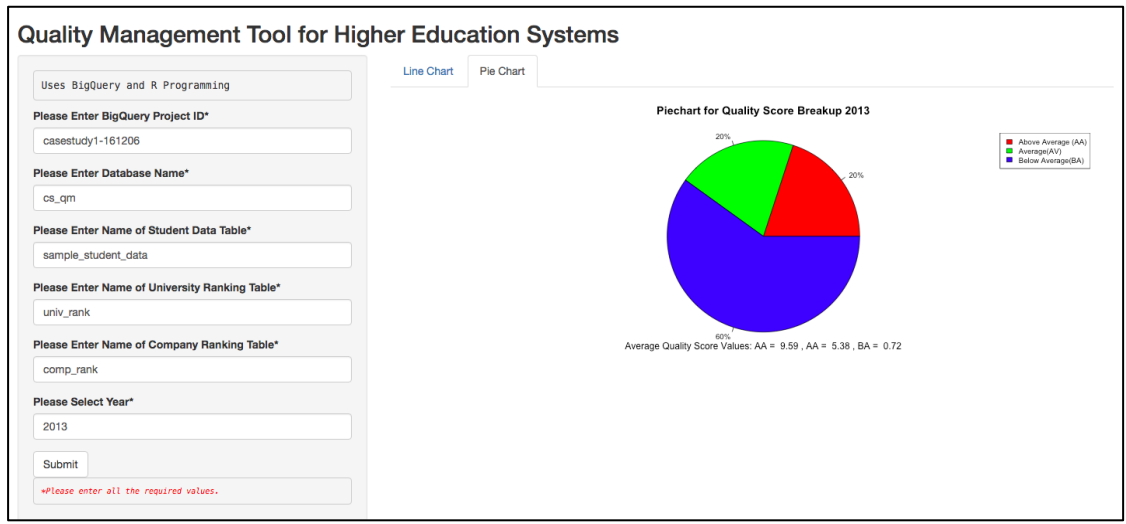

Figure 6 – PieChart Generated for Year 2013

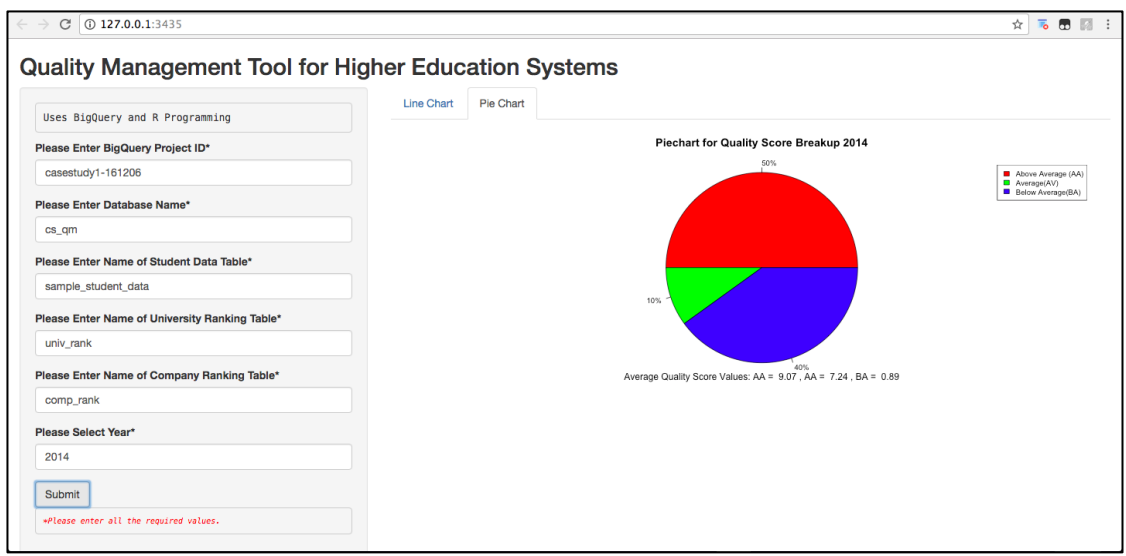

Figure 7 – PieChart Generated for Year 2014

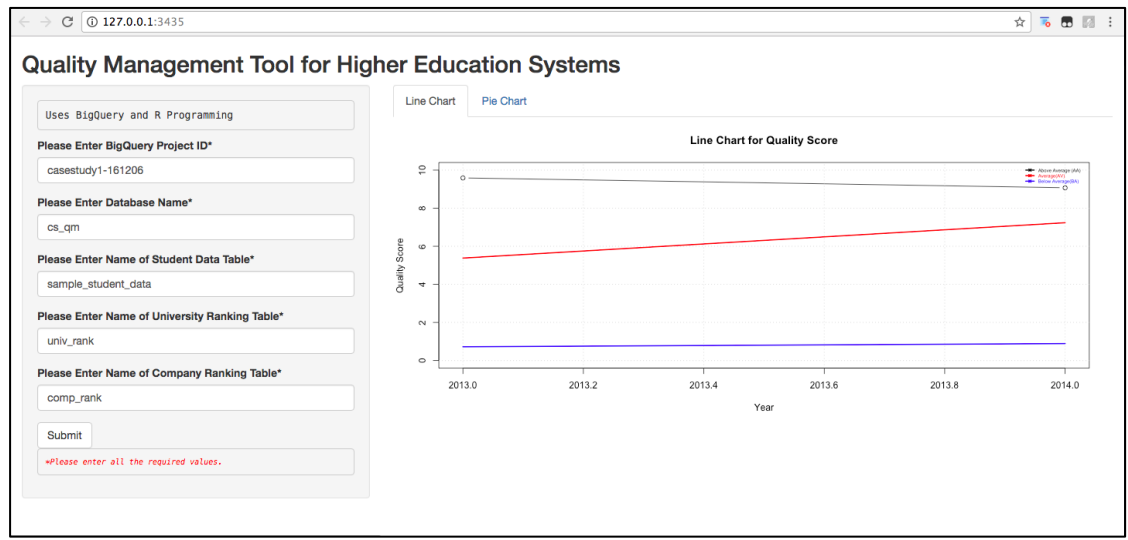

Figure 8 – Line Chart Generated

**Conclusion**

Existing quality frameworks proposed for higher education cover many dimensions of quality including content and delivery. However, most of these frameworks consider education as a service and evaluate quality from a service-oriented perspective. As a result, the fact that students are a product, besides being a consumer, is neglected.

The proposed framework uses this unexplored dimension of quality. In view of the fact that students are the outcomes produced by higher education institutions, an outcome-based analysis of this transition data can give useful insights into the quality of education provided by the university. Moreover, analytics of this nature can be helpful in decision-making and administrative planning for the educational institutions, contributing to quality improvement.

The framework is implemented in the form of a web-based application, using big data technologies that include Google BigQuery and R programming. It is tested using a dummy dataset to demonstrate how quality scores are calculated. With the help of the generated quality scores, analytical visualizations like piecharts and line charts are generated.

It is important to mention that the data available for analysis is limited to students who take up a postgraduate course/research degree or campus placement after the completion of their respective courses. Data about students, who get jobs through off-campus placement immediately after the degree or after taking a gap year, is usually not available. Moreover, data collection related to students who opt for self-employment is also limited. However, off-campus placements within one year of course completion are only considered because the framework generates yearly analysis.

This limits the capabilities of the proposed analytic framework by restricting the amount and diversity of data available for analysis. A more robust data collection at the organization level or alumni association levels can be a significant step towards a more efficient outcome-based analysis in the educational context. Moreover, self-employment is not accounted for in the framework.

Future research work shall include designing a comprehensive data collection framework and exploration of other variables that may affect and govern outcomes analysis other than transitions from one degree to the next level. In addition, academic categories like self-employed students can also be explored to extend the framework and make it more comprehensive.

## Acknowledgements

This work was supported by a grant from "Young Faculty Research Fellowship" under Visvesvaraya PhD Scheme for Electronics and IT, Department of Electronics & Information Technology (DeitY), Ministry of Communications & IT, Government of India.